\documentstyle[preprint,aps]{revtex}
\begin{document}
\draft
\title{Phases of Josephson Junction Ladders}
\author{Colin Denniston$^{1}$ and Chao Tang$^{2}$}
\address{$^{1}$Department of Physics, Princeton University,
Princeton, New Jersey 08544}
\address{$^{2}$NEC Research Institute, 4 Independence Way, Princeton,
New Jersey 08540}

\date{\today}
\maketitle

\begin{abstract}
We study a Josephson junction ladder in a magnetic field in the absence of
charging effects via a transfer matrix formalism.  The eigenvalues of the
transfer matrix are found numerically, giving a determination of the
different phases of the ladder.  The spatial periodicity of the ground state
exhibits a devil's staircase as a function of the magnetic flux filling factor
$f$.  If the transverse Josephson coupling is varied a continuous
superconducting-normal transition in the transverse direction is observed,
analogous to the breakdown of the KAM trajectories in dynamical systems.
\end{abstract}

\narrowtext

Two-dimensional arrays of Josephson junctions have attracted much recent
theoretical and experimental attention \cite{jja}.  Interesting physics
arises as a result of competing vortex-vortex and vortex-lattice
interactions.  It is also considered to be a convenient experimental
realization of the frustrated XY models.  In this paper, we discuss the
simplest such system, namely the Josephson junction ladder (JJL)
\cite{kardar,granato,mooij} shown in the inset of Fig.~\ref{stairvj}.

To construct the system, super\-conduct\-ing elements are placed at the ladder
sites.  Below the bulk super\-conducting-normal transition temperature,
the state of each element is described by its charge and the phase of the
superconducting wave function \cite{anderson}.  In this paper we neglect
charging effects, which corresponds to the condition that $4e^2/C \ll
J$, with $C$ being the capacitance of the element and $J$ the Josephson
coupling.  Let $\theta_j$ ($\theta_j'$) denote the phase on the
upper (lower) branch of the ladder at the $j$'th rung.  The Hamiltonian
for the array \cite{tinkham} can be written in terms the gauge invariant
phase differences, $\gamma_j=\theta_{j}-\theta_{j-1}-(2\pi/\phi_0)
\int_{j-1}^j A_x dx$, $\gamma_j'=\theta_{j}'-\theta_{j-1}'-(2\pi/\phi_0)
\int_{j'-1}^{j'} A_x dx$, and $\alpha_j=\theta_{j}'-\theta_{j}
-(2\pi/\phi_0) \int_{j}^{j'} A_y dx$:
\begin{equation}
{\cal H} = - \sum_j (J_x\cos\gamma_j + J_x\cos\gamma_{j}'
+ J_y\cos\alpha_j),
\label{ham1}
\end{equation}
where $A_x$ and $A_y$ are the components of the magnetic vector potential
along and transverse to the ladder, respectively, and $\phi_0$ the flux
quantum.  The sum of the phase
differences around a plaquette is constrained by
$\gamma_j-\gamma_{j}'+ \alpha_{j}-\alpha_{j-1}=2 \pi (f-n_{j})$, where
$n_j=0, \pm 1, \pm 2, ...$ is the vortex occupancy number and $f=\phi/\phi_0$
with $\phi$ being the magnetic flux through a plaquette.  With this
constraint, it is convenient to write Eq.~(\ref{ham1}) in the form
\begin{eqnarray}
{\cal H} &=&-J\sum_j \{2\cos\eta_j\cos[(\alpha_{j-1}-\alpha_{j})/2+
           \pi(f-n_j)]  \nonumber\\
         & & \quad\quad\quad \mbox{}+ J_t\cos\alpha_j\},
\label{ham2}
\end{eqnarray}
where $\eta_j=(\gamma_j+\gamma_j')/2$, $J=J_x$ and $J_t=J_y/J_x$.   The
Hamiltonian is symmetric under $f\rightarrow f+1$ with $n_j\rightarrow n_j+1$,
 and $f\rightarrow -f$ with $n_j\rightarrow -n_j$, thus it is sufficient to
study only the region $0 \le f \le 0.5$.  Since in one dimension ordered
phases occur only at zero temperature, the main interest is in the ground
states of the ladder and the low temperature excitations.
Note that in Eq.~(\ref{ham2}) $\eta_j$ decouples from $\alpha_j$ and $n_j$, so
that all the ground states have $\eta_j=0$ to minimize $\cal H$.
The ground states will be among the solutions to the current conservation
equations $\partial {\cal H}/\partial \alpha_j=0$:
\begin{eqnarray}
J_t\sin\alpha_j &=&
\sin[(\alpha_{j-1}-\alpha_{j})/2+\pi(f-n_j)] \nonumber\\
                 & & \mbox{} - \sin[(\alpha_j-\alpha_{j+1})/2+\pi(f-n_{j+1})].
\label{ccons}
\end{eqnarray}
For any given $f$ there are a host of solutions to Eq. (\ref{ccons}). The
solution that minimizes the energy must be selected to obtain the ground state.

If one expands the inter-plaq\-uette coup\-ling term in Eq.~(\ref{ham2}),
$\cos[(\alpha_{j-1}-\alpha_{j})/2 +\pi(f-n_j)]$, about it's max\-i\-mum, the
disc\-rete sine-Gordan model is obtained.
A vortex ($n_j=1$) in the JJL corresponds to a kink in the sine-Gordan model.
Kardar \cite{kardar} used this analogy to argue that this system should show
similar behavior to the discrete sine-Gordan model which has been studied by
several authors \cite{aubry,cop,Pokrovsky}. This analogy is only valid for
$J_t$ very small so that the inter-plaquette term dominates the behavior of the
system making the expansion about its maximum a reasonable assumption.
However, much of the interesting behavior of the discrete sine-Gordan model
occurs in regions of large $J_t$ ($J_t\sim 1$).  Furthermore, much of the work
by Aubry \cite{aubry} on the sine-Gordan model relies on the convexity of the
coupling potential which we do not have in the JJL.

In this Letter we formulate the problem in terms of a transfer matrix
obtained from the full partition function of the ladder.  The eigenvalues
and eigenfunctions of the transfer matrix are found numerically to
determine the phases of the ladder as functions of $f$ and $J_t$.  We find
that the spatial periodicity of the ground states goes through a devil's
staircase as a function of $f$.  We then study the properties of various
ground states and the low temperature excitations.  As $J_t$ is varied, all
incommensurate ground states undergo a superconducting-normal transition at
certain $J_t$ which depends on $f$.  One such transition will be analyzed.
Finally we discuss the critical current.

The partition function for the ladder, with periodic boundary conditions and
$K=J/k_BT$, is
\widetext
\begin{eqnarray}
{\cal Z} &=& \prod_i^N \int_{-\pi}^\pi \sum_{\{n_i\}} d\alpha_i d\eta_i
\exp\left\{K(2 \cos\eta_i \cos[(\alpha_{i-1}-\alpha_i)/2+\pi(f-n_i)]
+J_t\cos\alpha_i) \right\}.
\end{eqnarray}
The $\eta_i$ can be integrated out resulting in a simple transfer matrix
formalism for the partition function involving only the transverse phase
differences:
${\cal Z} =\prod_i^N \int_{-\pi}^\pi d\alpha_i P(\alpha_{i-1},\alpha_i)
=Tr\, \hat P^N.$
The transfer matrix elements $P(\alpha,\alpha')$ are
\begin{equation}
P(\alpha,\alpha') = 4\pi \exp[KJ_t(\cos\alpha+\cos\alpha')/2] \,
I_{0}(2 K \cos[(\alpha-\alpha')/2+\pi f]),
\label{mat}
\end{equation}
\narrowtext
where $I_0$ is the zeroth order modified Bessel function.  Note that
the elements of $\hat P$ are real and positive, so that its largest
eigenvalue $\lambda_0$ is real, positive and nondegenerate.  However, since
$\hat P$ is not symmetric (except for $f=0$ and $f=1/2$) other eigenvalues
can form complex conjugate pairs.  As we will see from the correlation
function, these complex eigenvalues determine the spatial periodicity of the
ground states.

The two point correlation function of $\alpha_j$'s is
\begin{eqnarray}
\langle e^{i(\alpha_0-\alpha_l)}\rangle & = & \lim_{N\rightarrow\infty}
\frac{ \left( \prod_{i}^N \int_{-\pi}^\pi d\alpha_{i}
P(\alpha_{i-1},\alpha_i) \right) e^{i(\alpha_0-\alpha_l)}}{\cal Z}
\nonumber \\
& = & \sum_n c_n \left(\frac{\lambda_n}{\lambda_0}\right)^l,
\label{cor}
\end{eqnarray}
where we have made use of the completeness of the left and right
eigenfunctions.  (Note that since $\hat P$ is not symmetric both right
$\psi_n^R$ and left $\psi_n^L$ eigenfunctions are need for the evaluation
of correlation functions.)  The $\lambda_n$ in Eq.~(\ref{cor}) are the
eigenvalues ($|\lambda_n|$ $\ge$ $|\lambda_{n+1}|$ and $n=0, 1, 2, ...$),
and the constants $c_n=\int_{-\pi}^\pi d\alpha'\psi_0^L(\alpha')
e^{i\alpha'}\psi_n^R(\alpha')\int_{-\pi}^\pi d\alpha\psi_n^L(\alpha)
e^{-i\alpha}\psi_0^R(\alpha)$.
In the case where $\lambda_1$ is real and $|\lambda_1|>|\lambda_2|$,
Eq.~(\ref{cor}) simplifies for large $l$ to
$$
\langle e^{i(\alpha_0-\alpha_l)}\rangle=c_0
                           +c_1\left(\frac{\lambda_1}{\lambda_0}\right)^l,
                           \quad |\lambda_1|>|\lambda_2|.
$$
In the case where $\lambda_1=\lambda_2^*=|\lambda_1|e^{i2\pi\Xi}$,
Eq.~(\ref{cor}) for large $l$ is
$$
\langle e^{i(\alpha_0-\alpha_l)}\rangle = c_0 +
  \left(c_1 e^{i2\pi\Xi l}+c_2 e^{-i2\pi\Xi l}\right)\left|\frac{\lambda_1}
 {\lambda_0}\right|^l , \, \lambda_1=\lambda_2^*.
$$
Note that while the correlation length is given by
$\xi=[\ln|\lambda_0/\lambda_1|]^{-1}$ the quantity $\Xi=Arg(\lambda_1)/2
\pi$ determines the spatial periodicity of the state.   Calculating $\lambda_n$
numerically \cite{colin}, we found that for $f$ smaller than a critical value
$f_{c1}$ which depends on $J_t$, both $\lambda_1$ and $\lambda_2$ are real.
These two eigenvalues become degenerate at $f_{c1}$, and then bifurcate into a
complex conjugate pair.  $\Xi$ as a function of $f$ is shown in
Fig.~\ref{stairvj} for several different values of $J_t$.  The shape of the
curves in Fig.~\ref{stairvj} is generally referred to as a devil's staircase.
The steps of the staircase are at $\Xi=p/q$, where $p$ and $q$ are integers.
These are commensurate states with $p$ vortices in each unit cell which
consists of $q$ plaquettes.  For small $J_t$, the flat steps are
connected by smooth differentiable curves; most states on the $\Xi-f$
curve are incommensurate states.  As $J_t$ increases,  more and more
steps appear and grow at the expense of the smooth regions.  At
$J_t=J_t^c\approx0.7$ the staircase becomes complete, i.e. there is a step
for every rational $\Xi$ and the set of $f$ which correspond to irrational
$\Xi$ has zero measure.  For $J_t>J_t^c$, the staircase becomes
over-complete, i.e. steps of lower order rationals grow and steps of
higher order rationals disappear \cite{foot1}.  Another important
characterization of a state is the phase density $\rho(\alpha)$:
$\rho(\alpha)d\alpha$ is the average fraction of all sites
in the ladder with $\alpha<\alpha_i<\alpha+d\alpha$.  If $\rho(\alpha)$
is a smooth function and $\rho(\alpha)>0$ for $\alpha \in (-\pi,\pi]$ at
$T=0$, the ground state energy is invariant under an adiabatic change of
$\alpha$'s.  Consequently, there is no phase coherence between upper and
lower branches of the ladder and hence no superconductivity in the
transverse direction.  In this case, we say that the $\alpha$'s are unpinned.
If there exist finite intervals of $\alpha$ on which $\rho(\alpha)=0$, there
will be phase coherence between the upper and lower branches and we
say that the $\alpha$'s are pinned.  In term of the transfer matrix, the
phase density is the product of the left and right eigenfunctions of
$\lambda_0$ \cite{foot2}, $\rho(\alpha)=\psi_0^L(\alpha) \psi_0^R(\alpha)$.

We first discuss the case where $f<f_{c1}$.  These are the ``Meissner''
states in the sense that there are no vortices ($n_i=0$) in the ladder.
The ground state is simply $\alpha_i=0$, $\gamma_j=\pi f$ and $\gamma_j'=-\pi
f$, so that there is a global screening current $\pm J_x\sin \pi f$ in the
upper and lower branches of the ladder \cite{kardar}.  The phase density
$\rho(\alpha)=\delta (\alpha)$.  Fig.~\ref{stairvj}(c) shows that at $J_t=1$,
the Meissner state extends all the way from $f=0$ to $f=f_{c1}\approx 0.28$.
The properties of the Meissner state can be studied by expanding
Eq.~(\ref{ham2}) around $\alpha_i=0$:
${\cal H}_M = (J/4)\sum_j [\cos(\pi f)(\alpha_{j-1}-\alpha_j)^2
+2 J_t \alpha_i^2]$.
The current conservation Eq.~(\ref{ccons}) becomes
\begin{equation}
\alpha_{j+1}=2 \left( 1+J_t/\cos \pi f\right)\alpha_j-\alpha_{j-1}.
\label{fin_diff}
\end{equation}
Besides the ground state $\alpha_j=0$, there are other two linearly
independent solutions $\alpha_j=e^{\pm j/\xi_M}$ of Eq.~(\ref{fin_diff})
which describe collective fluctuations about the ground state, where
\begin{equation}
{1 \over \xi_M}=\ln\left[1+{J_t \over \cos\pi f}+\sqrt{
    {2 J_t \over \cos\pi f}+\left({J_t \over \cos\pi f}\right)^2}\,\right].
\label{cor-len}
\end{equation}
$\xi_M$ is the low temperature correlation length for the Meissner state.
(Note that $\xi_M<1$ for $J_t\sim 1$ making a continuum approximation invalid.)
As $f$ increases, the Meissner state becomes unstable to the formation of
vortices.  A vortex is constructed by patching the two solutions of
Eq.~(\ref{fin_diff}) together using a matching condition.  The energy
$\epsilon_v$ of a single vortex is found to be
\begin{eqnarray}
\epsilon_v & \approx & [2+(\pi^2/8)\tanh(1/2\xi_M)]\cos \pi f \nonumber\\
           & & \mbox{} - (\pi+1)\sin\pi f+2 J_t,
\label{sol-e}
\end{eqnarray}
for $J_t$ close to one.  The zero of $\epsilon_v$ determines $f_{c1}$ which is
in good agreement with the numerical result from the transfer matrix.  For
$f>f_{c1}$, $\epsilon_v$ is negative and vortices are spontaneously created.
When vortices are far apart their interaction is caused only by the
exponentially small overlap.  The corresponding repulsion energy is of the
order $J\exp(-l/\xi_M)$, where $l$ is the distance between vortices.  This
leads to a free energy per plaquette of $F=\epsilon_v / l + J \exp(-l/\xi_M)
/ l$ \cite{Pokrovsky}.  Minimizing this free energy as a function of $l$ gives
the vortex density for $f>f_{c1}$: $\langle n_j \rangle=l^{-1}=
[\xi_M\ln|f_{c1}-f|]^{-1}$ where a linear approximation is used for $f$ close
to $f_{c1}$.

We now discuss the commensurate vortex states, taking the one with $\Xi=1/2$ as
an example.  This state has many similarities to the Meissner state but some
important differences.  The ground state is
\begin{eqnarray}
\alpha_0 & = & \arctan\left[ {2 \over J_t} \sin(\pi f)\right], \
\alpha_1=-\alpha_0, \ \alpha_{i\pm 2} = \alpha_i; \nonumber \\
n_0 & = & 0, \ n_1=1, \ n_{i\pm 2}=n_i,
\label{2state}
\end{eqnarray}
so that there is a global screening current in the upper and lower branches of
the ladder of $\pm 2\pi J (f-1/2)/\sqrt{4+J_t^2}$.  The existance of
the global screening, which is absent in an
infinite 2D array, is the key reason for the existance of the steps at
$\Xi=p/q$.  It is easy to see that the symmetry of this vortex
state is that of the (antiferromagnetic) Ising model.  The ground state is
two-fold degenerate.  The low temperature excitations are domain boundaries
between the two degenerate ground states.  The energy of the domain boundary
$J\epsilon_b$ can be estimated using similar methods to those used to derive
Eq.~(\ref{sol-e}) for the Meissner state.  We found that
$\epsilon_b = \epsilon_b^0 - (\pi^2/\sqrt{4+J_t^2})|f-1/2|$, where
$\epsilon_b^0$ depends only on $J_t$.
Thus the correlation length diverges with temperature as
$\xi\sim\exp(2J\epsilon_b /k_BT)$.  The transition from the $\Xi=1/2$ state
to nearby vortex states happens when $f$ is such that $\epsilon_b=0$; it is
similar to the transition from the Meissner state to its nearby vortex states.
 All other steps $\Xi=p/q$ can be analyzed similarly.  For comparison, we have
evaluated $\xi$ for various values of $f$ and $T$ from the transfer matrix
and found that $\xi$ fits $\xi\sim\exp(2 J\epsilon_b/k_BT)$ (typically over
several decades) at low temperature.  The value of $\epsilon_b$ as a function
of $f$ is shown in Fig.~\ref{tau} for $J_t=1$.  The agreement with the above
estimate for the $\Xi=1/2$ step is excellent. The tips of the peaks in
Fig.~\ref{tau} for states with $\Xi=1/q$ fit the relationship
$\tau\sim\exp(-q/l_0)$ with $l_0 \approx 0.77$, which is in good agreement with
$\xi_M(f=0)\approx0.76$ of Eq.~(\ref{cor-len}).


We now discuss the superconducting-normal transition in the transverse
direction.  For $J_t=0$, the ground state has $\gamma_i=\gamma_i'=0$ and
\begin{equation}
\alpha_j=2\pi f j+\alpha_0 -2 \pi \sum\nolimits_{i=0}^{i=j} n_i.
\label{j0state}
\end{equation}
The average vortex density $\langle n_j \rangle$ is $f$; there is no screening
of the magnetic field.  $\alpha_0$ in Eq.~(\ref{j0state}) is arbitrary; the
$\alpha$'s are unpinned for all $f$.  The system is simply two uncoupled 1D XY
chains, so that the correlation length $\xi=1/k_BT$.  The system
is superconducting at zero temperature along the ladder, but not in the
transverse direction.  As $J_t$ rises above zero we observe a distinct
difference between the system at rational and irrational values of $f$.  For
$f$ rational, the $\alpha$'s become pinned for $J_t>0$ ($\rho(\alpha)$ is a
finite sum of delta functions) and the ladder is superconducting in {\it both}
the longitudinal and transverse directions at zero temperature.  The behavior
for irrational $f$ is illustrated in the following for the state with
$\Xi=a_g$, where $a_g\approx 0.381966\cdots$ is one minus the inverse of the
golden mean.  Fig.~\ref{kam_den} displays $\rho(\alpha)$ for several different
$J_t$ at $\Xi=a_g$.  We see that the zero-frequency phonon mode (the smoothness
of $\rho(\alpha)$) persists for small $J_t>0$ until a critical value
$J_t^c(f)\approx 0.7$ where the $\alpha$'s
become pinned and the ladder becomes superconducting in the transverse
direction.  In the sine-Gordan model, the pinning transition of this
state coincides with the devil's staircase of Fig.~\ref{stairvj} becoming
complete \cite{aubry,cop} (If the $\alpha_j$'s are pinned in this state, then
all incommensurate states should be pinned).  The pinning transition of
the incommensurate states can be also studied using Eq.~(\ref{ccons}) which
can be viewed as a two-dimensional map.  The disappearance of the
zero-frequency phonon mode for irrational $\Xi$'s at finite small $J_t^c(f)$
is equivalent to the breakdown of the Kolmogorov-Arnold-Moser (KAM)
trajectories of the map \cite{KadShen}.

We now turn to the subject of critical currents along the ladder.  One can
obtain an estimate for the critical current by performing a perturbation
expansion around the ground state (i.e. $\{n_j\}$ remain fixed) and imposing
the current constraint of $\sin\gamma_j+\sin\gamma_j'=I$.  Let
$\delta\gamma_j$, $\delta\gamma_j'$ and $\delta\alpha_j$ be the change of
$\gamma_j$, $\gamma_j'$ and $\alpha_j$ in the current carrying state.
One finds that stability of the ground state requires that $\delta\alpha_j=0$,
and consequently $\delta\gamma_j=\delta\gamma_j'=I/2\cos\gamma_j$.
The critical current can be estimated by the requirement that the
$\gamma_j$ do not pass through $\pi/2$, which gives $I_c =
2(\pi/2 - \gamma_{\max})\cos\gamma_{\max}$, where $\gamma_{\max}=\max_j
(\gamma_j)$.  In all ground states we examined, commensurate and
incommensurate, we found that $\gamma_{\max} < \pi/2$, implying a
finite critical current for all $f$.

In conclusion, we have studied the equilibrium behavior of a Josephson junction
ladder in a magnetic field in the absence of charging effects.  The screening
current plays an important role in this system.  For $f<f_{c1}$, there is a
``Meissner state'' with no vortices.  For $f>f_{c1}$, the spatial periodicity
of the ground state climbs a devil's staircase as a function of $f$.  All
incommesurate states undergo a superconducting-normal transition in the
transverse direction as $J_t$ is increased, so that for $J_t>J_t^c\approx 0.7$
the ladder is superconducting in both the longitudinal and transverse
directions for all $f$.  The critical current along the
ladder is found to be finite for all $f$.  Finally, although in one dimension
there is no finite temperature phase transition and no true long range order,
our study showed that the correlation lengths in vortex states are extremely
long for reasonably low temperatures.  Thus it would be interesting to test
these results experimentally.  More extensive details of this system, including
current carrying states will be presented elsewhere along with a similar
analysis of the sine-Gordan model \cite{colin}.

We thank Sue Coppersmith, Xinsheng Ling, and Qian Niu for many valuable
discussions.

\begin{figure}
\caption{Period\-icity, $\Xi=Arg(\lambda_1)/2\pi$ ver\-sus $f$ for
$k_BT/J$ $=$ $0.005$ and, (a) $J_t$ $=$ $0.3$, (b) $J_t$ $=$ $0.7$,
(c) $J_t$ $=$ $1.0$.  Inset: The Josephson junction ladder is formed by the
arrangement of the super\-conducting islands.  The field ${\bf H}$ is out of
the page and the arrows indicate the direction of the gauge invariant phase
differences.}
\label{stairvj}
\end{figure}

\begin{figure}
\caption{Effective Ising coupling as a function of $f$ for $J_t=1$.  The
inset shows the statistical error for $2\epsilon_b$ in the fitting.}
\label{tau}
\end{figure}

\begin{figure}
\caption{$\rho(\alpha)=\psi_0^L(\alpha)\psi_0^R(\alpha)$ versus $\alpha$ at
$k_BT/J=0.004$ and $\Xi$ $=$ $0.381\,966\,011\cdots$, and for (a) $J_t=0.4$;
(b) $J_t=0.65$; (c) $J_t=0.7$; and (d) $J_t=0.9$.  Note the smaller scale
for the upper plots.}
\label{kam_den}
\end{figure}

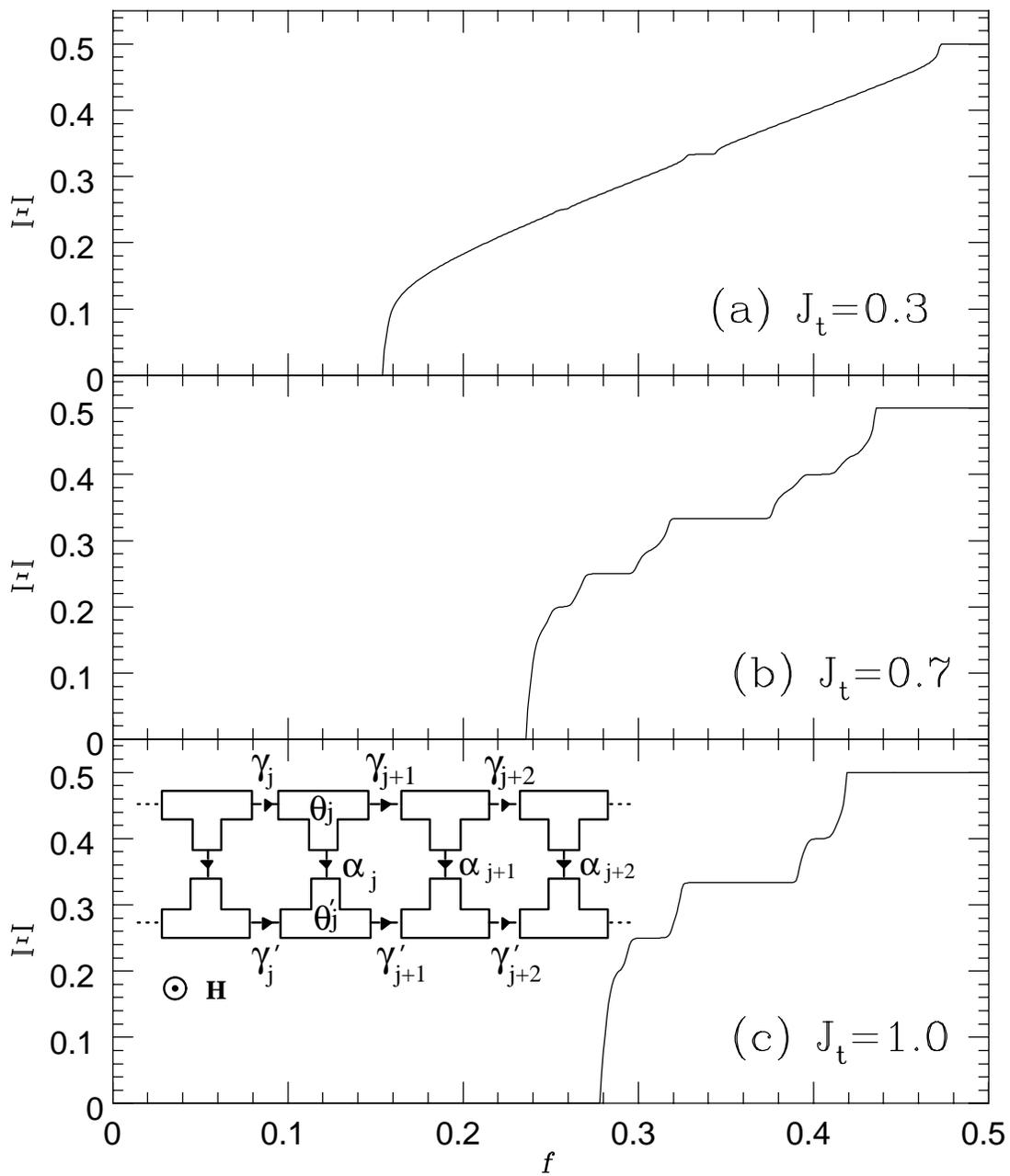
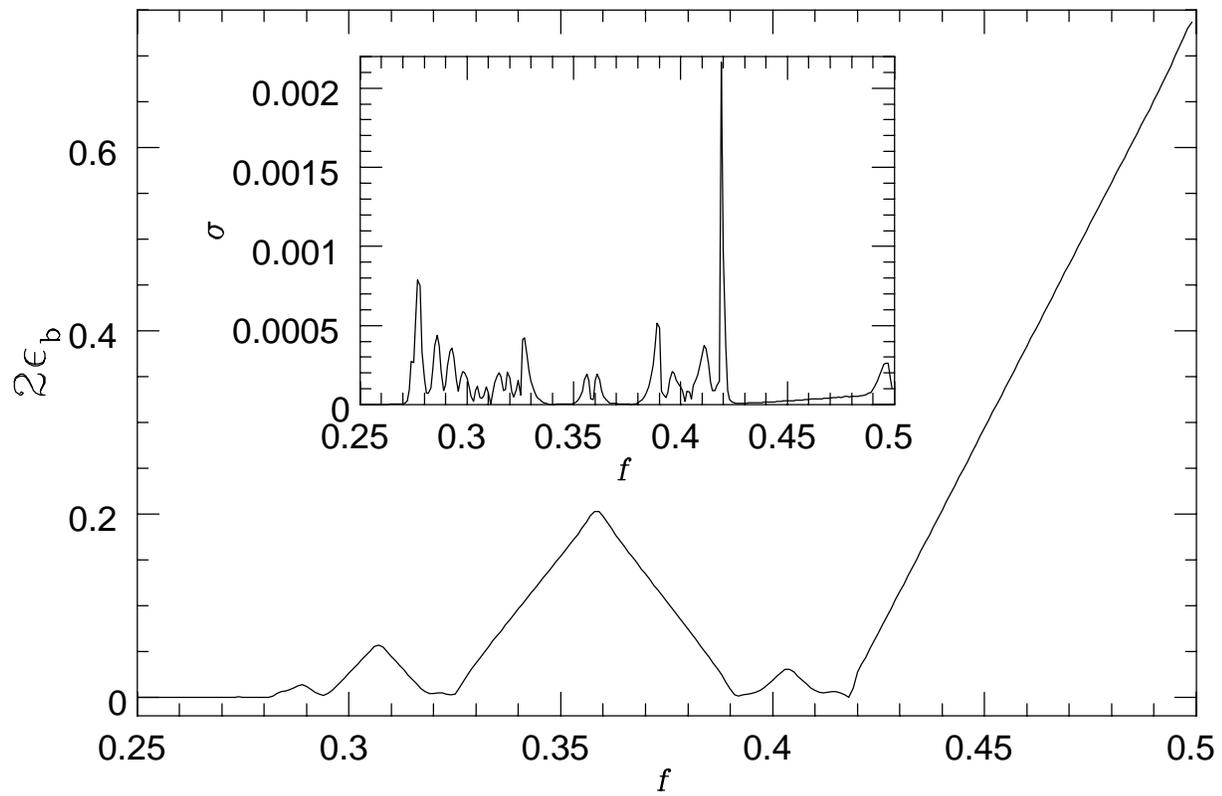
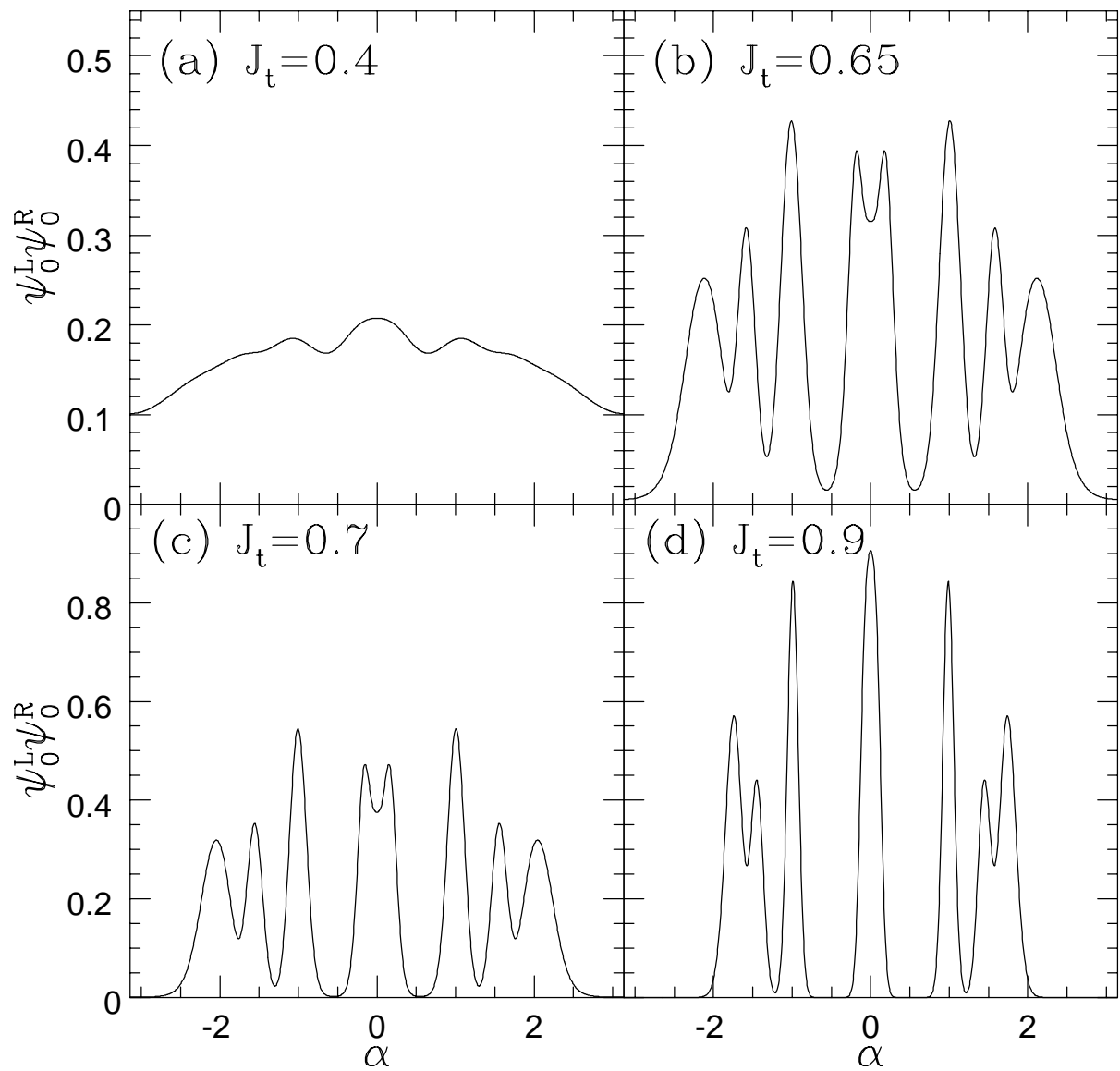
\end{document}